\documentclass[conference]{IEEEtran}
\IEEEoverridecommandlockouts
\newif\ifblind
\blindfalse  
\usepackage{cite}
\usepackage{amsmath,amssymb,amsfonts}
\usepackage{graphicx}
\usepackage{textcomp}
\usepackage{xcolor}
\usepackage{url}
\usepackage{xurl}
\usepackage{bm}
\usepackage{physics}
\usepackage[caption=false]{subfig}
\usepackage{stfloats}
\usepackage{verbatim}
\usepackage{booktabs}
\usepackage{hhline}
\usepackage[dvipsnames]{xcolor} 
\usepackage{pifont}             
\usepackage{wrapfig}
\usepackage{colortbl}
\usepackage{etoolbox}
\usepackage{array,tabularx}
\usepackage{placeins}
\usepackage{algorithm}
\usepackage{algpseudocode}
\usepackage{enumitem}
\usepackage{makecell}
\usepackage{listings}
\usepackage{xspace}

\usepackage[colorlinks=true,
            linkcolor=black,
            citecolor=black,
            urlcolor=blue]{hyperref}
\usepackage[normalem]{ulem}

\definecolor{LightGray}{gray}{0.92}
\definecolor{TIISPureBlue}{RGB}{0,0,255} 

\definecolor{oraKeyword}{RGB}{0,0,160} 
\definecolor{oraType}{RGB}{128,0,128} 
\definecolor{oraString}{RGB}{153,0,0} 
\definecolor{oraComment}{RGB}{0,128,0} 
\definecolor{oraNumber}{RGB}{96,96,96} 
\definecolor{oraRule}{RGB}{200,200,200} 

\newcolumntype{Y}{>{\centering\arraybackslash}X}
\newcolumntype{C}[1]{>{\centering\arraybackslash}m{#1}} 
\newcolumntype{L}[1]{>{\centering\arraybackslash}m{#1}} 
\newcolumntype{R}[1]{>{\centering\arraybackslash}m{#1}} 

\newcommand{\sys}{Miffie\xspace}

\newcommand{\appref}[1]{\hyperref[#1]{Appendix~\ref*{#1}}}
\newcommand{\figref}[1]{\hyperref[#1]{Fig.~\ref*{#1}}}
\newcommand{\tabref}[1]{\hyperref[#1]{Table~\ref*{#1}}}

\def\BibTeX{{\rm B\kern-.05em{\sc i\kern-.025em b}\kern-.08em
    T\kern-.1667em\lower.7ex\hbox{E}\kern-.125emX}}

\lstdefinestyle{sqlstyle}{
  language=SQL,
  basicstyle=\ttfamily\footnotesize,
  columns=fullflexible,
  breaklines=true,
  frame=single,
  rulecolor=\color{oraRule},
  keepspaces=true,
  showstringspaces=false,
  keywordstyle=\bfseries\color{oraKeyword},
  commentstyle=\itshape\color{oraComment},
  stringstyle=\color{oraString},
  morekeywords={
CREATE,TABLE,PRIMARY,KEY,FOREIGN,REFERENCES,CONSTRAINT,NOT,NULL,UNIQUE,DEFAULT,CHECK,INDEX,VIEW,SEQUENCE,TRIGGER,INSERT,INTO,VALUES,UPDATE,SET,DELETE,FROM,WHERE,JOIN,LEFT,RIGHT,INNER,OUTER,ON, GROUP,BY,ORDER,HAVING,DISTINCT,AS,AND,OR,IN,EXISTS,LIKE,ALTER,ADD,DROP,MODIFY
  },
  emph={INT,INTEGER,NUMBER,FLOAT,DOUBLE,DECIMAL,NUMERIC,
        VARCHAR,VARCHAR2,CHAR,NCHAR,NVARCHAR2,TEXT,
        DATE,DATETIME,TIMESTAMP},
  emphstyle=\bfseries\color{oraType}
}

\begin{document}

\title{Database Normalization via Dual-LLM Self-Refinement}

\ifblind
    \author{\IEEEauthorblockN{Anonymous Author(s)}
    \IEEEauthorblockA{\textit{Anonymous Affiliations}}}
\else
    \author{\IEEEauthorblockN{Eunjae Jo, Nakyung Lee, and Gyuyeong Kim*}
    \IEEEauthorblockA{\textit{Department of Computer Science} \\
    \textit{Sungshin Women's University}\\
    Seoul, South Korea \\
    \{220256023, 220254009, gykim\}@sungshin.ac.kr}
    \IEEEauthorblockA{*Corresponding author}
    }
\fi

\maketitle

\begin{abstract}
Database normalization is crucial to preserving data integrity.  However, it is time-consuming and error-prone, as it is typically performed manually by data engineers who must identify functional dependencies and systematically decompose schemas to satisfy multiple normal forms. To this end, we present Miffie, a database normalization framework that leverages the reasoning and code-generation capabilities of large language models. Miffie enables automated data normalization with minimal human effort while preserving high accuracy across 1NF through BCNF compliance. The core of Miffie is a dual-model self-refinement architecture that combines the best-performing models for normalized schema generation and verification, respectively. The generation module eliminates anomalies based on the feedback of the verification module until the output schema satisfies the requirement for normalization. We also carefully design task-specific zero-shot prompts to guide the models for achieving both high accuracy and cost efficiency. Experimental results show that Miffie can normalize complex database schemas while maintaining high accuracy. 
\end{abstract}

\begin{IEEEkeywords}
Relational databases, Data management, Large language models, Normalization, Self-refinement
\end{IEEEkeywords}

\section{Introduction \label{introduction}}
The rapid growth in the volume of data from web services has heightened the importance of maintaining data integrity in relational databases.
Normalization is a key to preserve data integrity~\cite{normal87,normal91} by following a set of normal forms (e.g., 1NF, 2NF, and 3NF), each of which addresses issues within relational schemas, such as removing non-atomic columns and functional dependencies.
Unfortunately, normalization remains an expert-driven task, typically performed manually by data engineers.
This is because normalization involves understanding domain-specific data semantics and context, which are hard to automate.
As datasets grow in size, normalization becomes increasingly time-consuming and error-prone, calling for an efficient mechanism to reduce human effort.
Although there are many data management tools~\cite{mysqlworkbench,oracleapex,griffith_norm_tool,openrefine,hayashi_normalizer,uis_db_design_tool,autonormalize,sqldeveloper}, they do not provide functionality to reorganize database schemas to satisfy the requirements of normal forms.

Meanwhile, recent advances in large language models (LLMs) have opened up opportunities for automated normalization thanks to their symbolic-reasoning capabilities~\cite{chainofthoughts,tominllm,attention}.
For example, LLMs can interpret structured data and detect violations of functional dependencies quickly.
However, simply applying LLMs to database normalization with na\"{\i}ve prompts is not enough because the generated results may be inaccurate due to nuanced semantic relationships between columns, which are difficult to capture.
In this context, we ask the following question: \textit{how can we automate database normalization while ensuring high accuracy?}

This paper answers the question by presenting \sys, a LLM-based database normalization framework.
The core of \sys is a dual-model self-refinement architecture that enables accurate and automated database normalization.
The self-refinement~\cite{selfrefine} is a general approach where a language model refines generated outputs iteratively based on the feedback from itself.
Unlike the original approach, our dual-model architecture uses different language models for the generation and feedback phases to optimize the database normalization process.
Furthermore, we carefully design task-specific zero-shot prompts~\cite{zeroshot_crs,llm4msr,twoheads} to guide the models to achieve high accuracy and cost efficiency simultaneously, which is also different from the original approach that uses cost-inefficient few-shot prompting~\cite{fewshot}.

The \sys framework operates through two primary components: a generation module and a verification module.
Initially, the generation module produces a normalized schema derived from the user's input schema.
Subsequently, the verification module rigorously assesses the correctness of this output.
If the schema fails to meet normalization standards, the verification module generates evaluation feedback, which the generation module uses to refine the schema.
This iterative refinement process continues until the verification module confirms the schema's full normalization.

To evaluate \sys, we employ a set of practical database schemas with varying levels of complexity, drawn from real-world domains such as online advertising, airport management, and e-commerce order processing.
We investigate the capability of \sys to normalize schemas containing anomalies across diverse normal forms within these practical settings.
Our results demonstrate that \sys detects anomalies with both speed and high accuracy, proving effective even for complex, real-world schemas.
Furthermore, we show that our task-specific zero-shot prompts achieve accuracy comparable to, or exceeding, that of few-shot prompts while minimizing token consumption.

In summary, our contributions are as follows.
\begin{itemize}[noitemsep]
\item{
To the best of our knowledge, \sys is the first LLM-based database normalization framework that significantly reduces human effort to preserve data integrity while maintaining high accuracy.
}
\item{
We propose a dual-model self-refinement architecture with task-specific zero-shot prompting that enables two models to collaborate to generate accurate normalized schemas via generation and verification loops.
}
\item{
We conduct a series of comprehensive experiments to demonstrate the efficiency and robustness of the \sys framework.
}
\end{itemize}

 
\section{Related Work \label{relatedwork}}
We briefly review existing work on traditional data management tools and recent LLM-based approaches for data tasks, both of which offer only limited automation for schema normalization.
In contrast, \sys differentiates itself by providing fully automated normalization through a novel dual-module refinement architecture.

\noindent~\textbf{Data management tool support for schema design.}
Several commercial and academic data management tools offer support for schema design and refinement~\cite{mysqlworkbench,oracleapex,openrefine,sqldeveloper,griffith_norm_tool,hayashi_normalizer,uis_db_design_tool}.
While some of these tools assist in normalization, they typically rely on a semi-automated process that requires users to manually specify functional dependencies (FDs).
Although helpful in ideally structured environments, this reliance proves problematic for real-world schemas where dependencies are often unclear, implicit, or incomplete.
Consequently, these tools struggle to scale effectively to complex scenarios and continue to demand extensive human intervention to guarantee correctness.

\noindent~\textbf{LLMs for data management.}
Recent research explores the application of LLMs to various data management tasks, primarily focusing on cleaning or formatting tabular content rather than structural reorganization.
Existing approaches leverage the semantic understanding of LLMs but stop short of the structural analysis required for rigorous database normalization.
For instance, Magneto~\cite{magneto} facilitates schema integration by aligning semantically related attributes across tables, utilizing LLMs to assess column matches retrieved by embedding models. 
Similarly, CHORUS~\cite{chorus} enhances table exploration by analyzing columns with an LLM-based model trained to classify semantic types, predict join keys, and identify primary entities. 
Finally, NormTab~\cite{normtab} employs LLMs to detect and rewrite inconsistent tabular values to improve interpretability, yet it does not address the decomposition of tables necessary for satisfying database normal forms.

\begin{figure*}[t]
\centering
\includegraphics[width=0.8\textwidth]{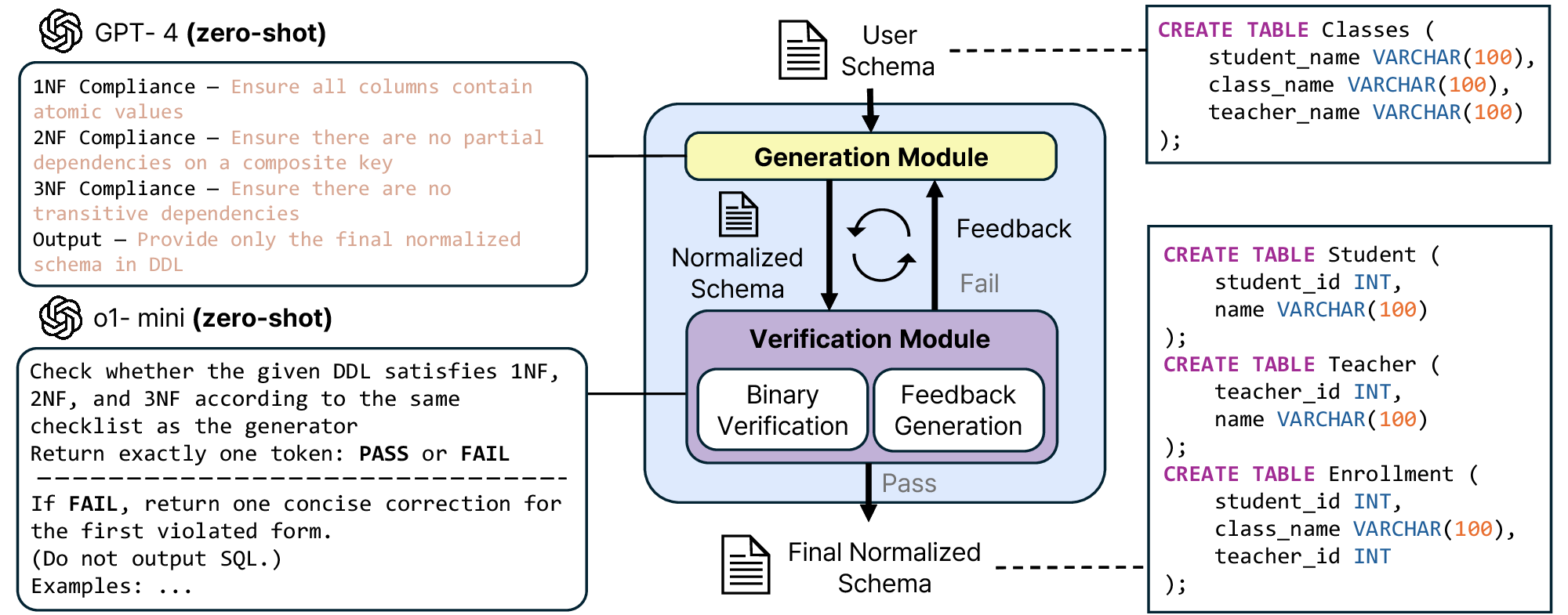}
\caption{The overview of \sys framework.
\label{fig:overview}}
\end{figure*}
\section{Design \label{design}}

This section presents the design of \sys and its core architectural components. 
We first provide an overview of the \sys framework and describe its iterative workflow. 
We then detail the dual-model self-refinement architecture and task-specific prompting strategy.

\subsection{\sys Framework}
Our goal is to automate database normalization while preserving high accuracy.
To achieve the goal, we design the \sys framework as shown in Fig.~\ref{fig:overview}.
\sys is based on our proposed dual-model self-refinement architecture.
The architecture consists of the generation module and the verification module with task-specific zero-shot prompts.
We use GPT-4 and o1-mini for the generation module and the verification module, respectively, by considering their overall performance in each functionality.

\noindent~\textbf{How it works.}
A user provides an initial schema as input to the framework.
The generation module produces a normalized schema as output.
The verification module checks this output to determine whether it satisfies the required normalization criteria.
If the verification module finds any violations of normalization requirements, it generates feedback including an explanation of the detected anomalies and instructions for resolving them.
Based on this feedback, the generation module refines the schema accordingly.
This process of schema generation and verification repeats iteratively until the verification module approves that the output schema indeed satisfies all normal form requirements or until a generation threshold (maximum number of attempts) is reached.

\begin{table}[t!]
\centering
\caption{Normalization accuracy (mean $\pm$ std) across different LLMs. The accuracy is the average number of removed anomalies. GPT-4 generally shows balanced accuracy.
\label{tab:acc_model-specific}}
\begin{tabular}{lccc} \toprule
Model             
& 1NF & 2NF & 3NF \\
\midrule
GPT-3.5-Turbo & 1.80 (\(\pm0.00\)) & 0.00 (\(\pm0.00\)) & 1.05 (\(\pm0.97\)) \\
GPT-4 & 4.30 (\(\pm0.66\)) & 4.70 (\(\pm0.71\)) & 4.35 (\(\pm0.91\)) \\
GPT-4-Turbo & 4.90 (\(\pm0.30\)) & 2.90 (\(\pm1.64\)) & 4.80 (\(\pm0.69\)) \\
GPT-4o-mini & 4.45 (\(\pm0.59\)) & 2.80 (\(\pm0.68\)) & 4.05 (\(\pm0.67\)) \\
o1-mini  & 4.80 (\(\pm0.40\)) & 3.30 (\(\pm0.47\)) & 4.45 (\(\pm0.50\)) \\
\bottomrule
\end{tabular}
\end{table}

\subsection{Dual-Model Self-Refinement Architecture}
The dual-model self-refinement is the core of \sys.
The general-purpose self-refinement approach~\cite{evolve} specifies that a single model refines the output based on the feedback from the same model.
In \sys, to maximize the efficiency in database normalization, we leverage the strengths of two different LLMs for the schema generation and verification, improving the accuracy of each task.
This is based on our observation that each LLM has different capabilities for the generation and verification tasks.

The generation module normalizes the given input schema, while the verification module evaluates the generated schema.
The verification module performs a binary verification for normal forms.
If an anomaly for any normal form is detected, it flags the schema as invalid for that normal form and all higher forms.
Next, it generates feedback that explains the detected anomaly and suggests detailed actions to resolve it, such as splitting tables.

\noindent~\textbf{Experiment 1: Finding the best model for generation.}
To identify the best-performing LLM for schema generation, we conduct a series of experiments.
Table~\ref{tab:acc_model-specific} shows the normalization accuracy of different LLMs for different normal forms.
We inject five anomalies for each normal form.
The accuracy here is defined as the average number of removed anomalies for 20 runs.
We can see that GPT-4 is the only model that stably removes anomalies across all the normal forms without performance variability.
The balanced accuracy of GPT-4 makes us to employ it for schema generation.
The other models do not have consistent performance. 
For example, GPT-4 Turbo is effective in removing anomalies for 1NF and 3NF, but it does not detect anomalies in 2NF well.
GPT-3.5 Turbo exhibits significantly lower accuracy across all normal forms, removing only a few anomalies on average.
Unless otherwise noted, we use the OpenAI API default sampling settings (temperature = 1.0) to avoid distortion caused by model-selection results with additional hyperparameter tuning.

\begin{figure}[t!]
\centering
\includegraphics[width=7.0cm]{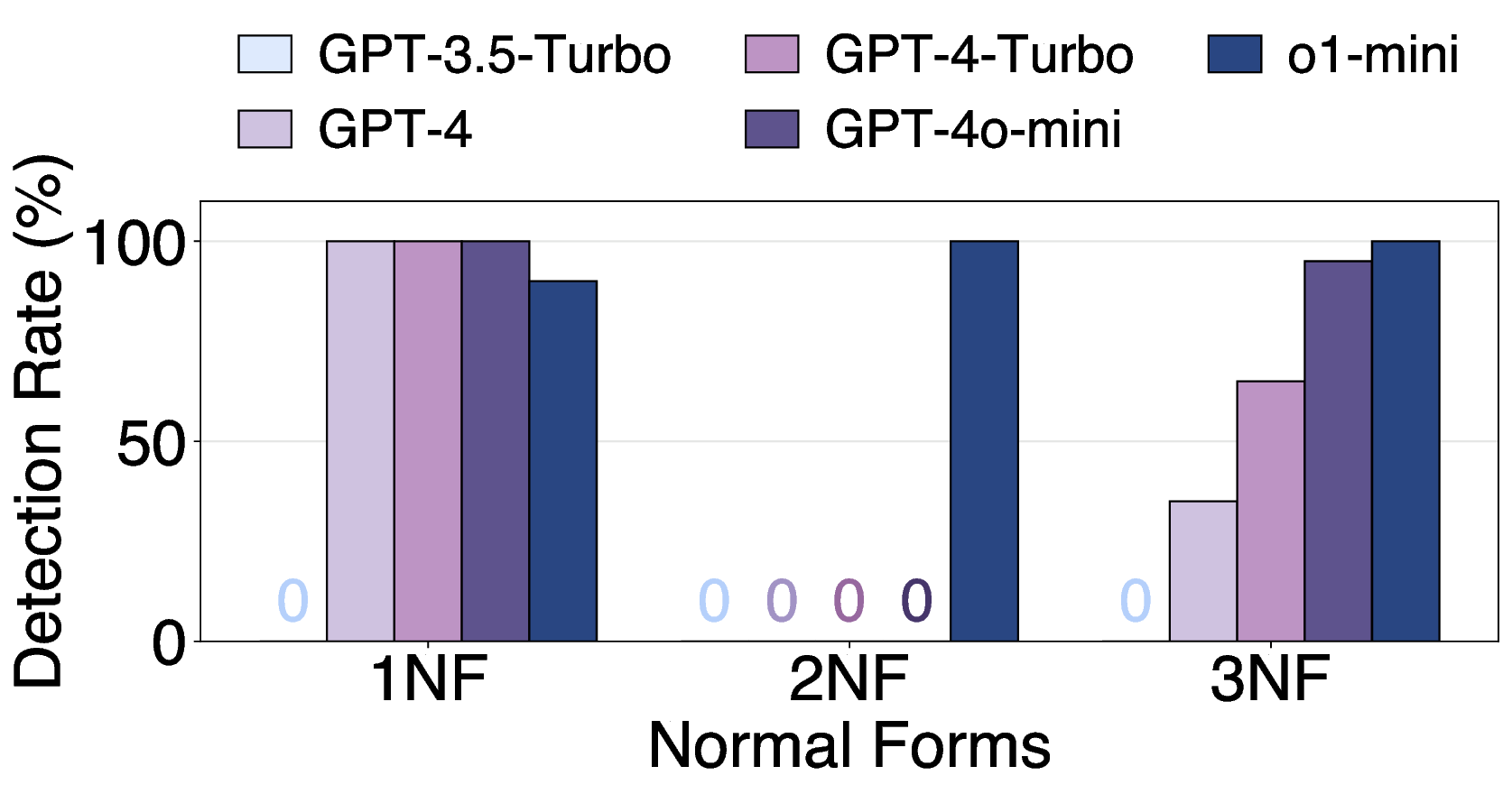}
\caption{Anomaly detection rates across different LLMs. OpenAI o1-mini constantly achieves high detection rates across all normal forms.}
\label{fig:verification_model-specific}
\end{figure}

\noindent~\textbf{Experiment 2: Finding the best model for verification.}
Fig.~\ref{fig:verification_model-specific} shows the anomaly detection rates of different models across the three normal forms.
The detection rate is defined as the number of detected anomalies divided by the five injected anomalies.
We observe that o1-mini achieves near-perfect detection rates across all normal forms with high consistency.
The other models show inconsistent results.
For example, while they can detect anomalies in 1NF, they fail to capture anomalies in 2NF.
Notably, GPT-3.5 Turbo fails to detect anomalies across all normal forms.

\begin{table}[t!]
\centering
\caption{Characteristic of Target Schemas.}
\label{tab:datasets}
\begin{tabular}{lccc}
\toprule
Schemas            
      & \makecell{\# of Tables}
      & \makecell{\# of Foreign Keys}
      & Complexity \\
\midrule
Orders~\cite{sqlallinone} &  4               &  3            & Easy       \\
Advertising~\cite{databaseanswers}   &  7               &  8            & Medium     \\
AirportDB~\cite{mysqldocument} & 14               & 21            & Hard       \\
\bottomrule
\end{tabular}
\end{table}
\subsection{Task-Specific Zero-Shot Prompting}
A naïve zero-shot prompt often fails to produce correctly normalized schemas due to ambiguous output formats and inconsistent interpretation of normalization criteria. 
To address this, we design task-specific prompts with structured instructions and fixed output format specifications. 
Notably, instead of providing complete input-output examples (i.e., few-shot), we explicitly list normalization criteria and specific violation patterns to guide the model.

\noindent~\textbf{Generation prompt.}
We design the generation prompt to instruct the model to normalize the input schema to satisfy the required normal forms and to output only the complete SQL code for the normalized schema, as shown in Fig.~\ref{fig:generation_prompt} in the Appendix. To stabilize generation and simplify automated parsing, we enforce a fixed output structure and explicitly prohibit interleaving descriptions or explanations with SQL statements. The prompt lists specific criteria and representative violations for each normal form to guide the model. For 1NF, we specify that all columns must contain atomic values, prohibiting multivalued attributes such as phone numbers stored as comma-separated lists. For 2NF, we provide simple examples of partial dependencies on composite keys, where a non-key attribute depends on only part of a composite primary key. For 3NF, we include a transitive dependency example $EmployeeID \rightarrow DeptID \rightarrow DeptName$, where $DeptName$ depends on a non-key attribute. 
By explicitly specifying these criteria in a structured way, we avoid relying on lengthy few-shot examples while still providing sufficient guidance to minimize formatting differences and normalization errors.

\noindent~\textbf{Verification prompt.}
The verification prompt specifies binary compliance checks for each normal form, assigning either 1 (compliant) or 0 (violation) to each normalization stage (see Fig.~\ref{fig:verification_prompt} in the Appendix). Following the monotonic nature of normalization, if the schema violates a criterion at a given stage, the verifier assigns 0 to that stage and all subsequent stages. In addition to the binary outcomes, the verifier returns a brief explanation of the detected anomaly, including violation details. This structured feedback directly guides the next refinement attempt by the generation module, enabling efficient iterative self-refinement until the schema is approved or a maximum attempt threshold is reached.

\begin{table}[t!]
\centering
\caption{Characteristic of Target Schemas.}
\label{tab:datasets}
\begin{tabular}{lccc}
\toprule
Schemas            
      & \makecell{\# of Tables}
      & \makecell{\# of Foreign Keys}
      & Complexity \\
\midrule
Orders~\cite{sqlallinone} &  4               &  3            & Easy       \\
Advertising~\cite{databaseanswers}   &  7               &  8            & Medium     \\
AirportDB~\cite{mysqldocument} & 14               & 21            & Hard       \\
\bottomrule
\end{tabular}
\end{table}
\section{Evaluation \label{evaluation}}

This section presents the experimental setup and evaluation results of \sys. 
We first describe the datasets, metrics, and baselines used in our experiments, and then analyze the impact of prompting strategies and architectural design choices. 
Finally, we examine the effects of refinement loops and schema complexity on normalization accuracy.

\subsection{Methodology}
\noindent~\textbf{Datasets and Benchmark Suite.}
Instead of relying on arbitrary real-world schemas that lack ground truth for normalization anomalies, we constructed a controlled benchmark suite to rigorously evaluate the logical reasoning capabilities of \sys.
Our suite consists of three representative database schemas from diverse domains~\cite{sqlallinone,databaseanswers,mysqldocument}: \textit{Orders} (E-commerce), \textit{Advertising} (Online Services), and \textit{AirportDB} (Large-scale Logistics), as detailed in Table~\ref{tab:datasets}.
These schemas were specifically chosen to cover a wide spectrum of structural complexity, ranging from simple transactional records to complex operational systems with intricate foreign key dependencies.

\noindent~\textbf{Controlled Anomaly Injection.}
Validating normalization requires a precise ground truth to measure whether specific functional dependencies are correctly resolved. Since "wild" legacy databases often contain ambiguous design choices rather than clear-cut normalization errors, we employed a systematic anomaly injection protocol.
We injected five distinct anomalies for each normal form (1NF, 2NF, 3NF) into the target schemas, creating a deterministic evaluation environment.
For \textit{Orders}, we introduced synthetic relations to facilitate 2NF and 3NF violations, ensuring that our evaluation covers all edge cases.
This approach allows us to conduct a total of 180 evaluation episodes (3 Schemas $\times$ 3 Normal Forms $\times$ 20 Trials), providing statistically significant metrics on the model's precision and recall.

\noindent~\textbf{Evaluation metrics.}
We use accuracy as the primary metric, defined as the number of correctly eliminated anomalies out of the five injected anomalies per normal form, averaged over 20 independent trials.
We also report the detection rate as the proportion corresponding to the average accuracy.

\noindent~\textbf{Baselines.} 
The vanilla refers to a naive zero-shot prompt that uses an unstructured instruction without providing any normalization criteria.
The zero-shot prompt uses detailed instruction that enumerates normalization criteria for 1NF-3NF and explicit examples of anomalies, but it does not use iterative feedback.
The single-model self-refine implements the iterative refinement loop using a single LLM to handle both schema generation and verification tasks. 
Our work, \sys, uses the dual-model self-refinement architecture that separates schema generation and verification tasks into GPT-4 and o1-mini, respectively, while iteratively refining the schema based on feedback.

\begin{figure}[t!]
\centering
\includegraphics[width=7.0cm]
{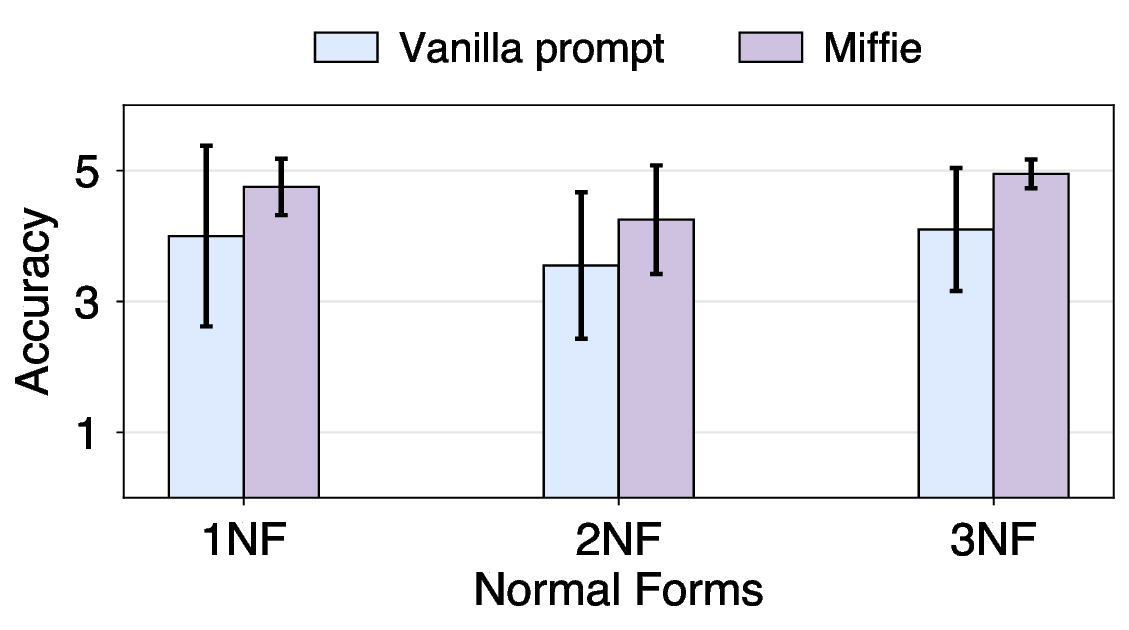}
\caption{Comparison of normalization accuracy between the vanilla prompt and \sys for each normal form.
\label{fig:mainResult}}
\end{figure}

\begin{table}[t]
\centering
\caption{Accuracy (mean $\pm$ std) and token usage under different prompting.
}
\label{tab:few-shot}
\begin{tabular}{lcccc}
\toprule
Prompts      
& 1NF
& 2NF
& 3NF 
& Tokens \\
\midrule
Zero-shot
  & 4.60 (\(\pm0.49\))
  & 4.10 (\(\pm1.51\))
  & 4.90 (\(\pm0.30\))
  & 325 \\
One-shot
  & 4.30 (\(\pm0.46\))
  & 4.20 (\(\pm0.98\))
  & 4.80 (\(\pm0.40\))  
  & 628 \\
Few-shot
  & 4.30 (\(\pm0.46\))
  & 4.80 (\(\pm0.40\))
  & 4.50 (\(\pm0.67\))
  & 1.1K \\
\bottomrule
\end{tabular}
\end{table}

\subsection{Results}
\noindent~\textbf{Overall comparison.} 
Fig.~\ref{fig:mainResult} shows the normalization accuracy of the vanilla prompt and \sys across the three normal forms.
\sys achieves higher accuracy than the vanilla prompt; the improvement is roughly 1.2× across all normal forms.
The vanilla prompt exhibits lower accuracy with variability because it uses unstructured instructions without explicit normalization criteria.
This result demonstrates that providing detailed instructions and using iterative feedback can improve accuracy.

\noindent~\textbf{Impact of prompting.}
We evaluate the impact of three different prompting: zero-shot, one-shot, and few-shot prompting.
The zero-shot prompt is the prompt used in \sys that specifies requirements for each normal form.
The one-shot prompt includes a single example of requirement violations for all normal forms.
The few-shot prompt contains examples of requirement violations for each normal form with the three target schemas.

Table~\ref{tab:few-shot} shows the results.
They indicate that the zero-shot prompt achieves comparable performance to the other prompting strategies across the normal forms while maintaining the best cost efficiency.
While the other prompts improve the accuracy in 2NF, the token usage is too large compared to the zero-shot prompt.
This result demonstrates that a carefully designed zero-shot prompt can achieve similar or even better performance with cost efficiency.

\begin{table}[t]
\centering
\caption{Accuracy comparison between single- and dual-model self-refinement architectures. 
}
\label{tab:single-dual}
\begin{tabular}{lcccc}
\toprule
Architecture          
& 1NF
& 2NF
& 3NF 
& \makecell{Elimination Rate\\($\le$3 tries)} \\
\midrule
\makecell{Single-model \\(GPT-4 only)}
  & \makecell{4.00 \\(\(\pm0.89\))}
  & \makecell{3.90 \\(\(\pm1.26\))}
  & \makecell{4.60 \\(\(\pm0.73\))}
  & 45\% \\
\makecell{Single-model \\(o1-mini only)}
  & \makecell{4.90 \\(\(\pm0.30\))}
  & \makecell{3.35 \\(\(\pm0.48\))}
  & \makecell{4.80 \\(\(\pm0.40\)) }
  & 57\% \\
\makecell{\textbf{Dual-model} \\ \textbf{(\sys)}}
  & \makecell{\textbf{4.75} \\(\(\pm0.43\))}
  & \makecell{\textbf{4.25} \\(\(\pm0.83\))}
  & \makecell{\textbf{4.95} \\(\(\pm0.22\))}
  & 72\% \\
\bottomrule
\end{tabular}
\end{table}

\begin{table}[t]
\centering
\caption{Impact of the verification module on normalization accuracy (mean $\pm$ std). 
}
\label{tab:prompt-compare}
\begin{tabular}{lccc}
\toprule
Method          
& 1NF
& 2NF
& 3NF \\
\midrule
w/o verification
  & 4.10 (\(\pm0.54\))
  & 4.10 (\(\pm1.48\))
  & 4.30 (\(\pm0.95\)) \\
\makecell{\textbf{w/ verification} \\ \textbf{(\sys)}}
  & \textbf{4.75} (\(\pm0.43\))
  & \textbf{4.25} (\(\pm0.83\))
  & \textbf{4.95} (\(\pm0.22\)) \\
\bottomrule
\end{tabular}
\end{table}

\noindent~\textbf{Impact of the number and type of models in self-refinement.}
In this experiment, we compare \sys with the single-model architectures to show that our dual-model architecture has better accuracy and detection rate.
Table~\ref{tab:single-dual} shows the accuracy and the elimination rate.
The elimination rate indicates the portion of cases when the architecture eliminates all the anomalies completely.
We can see that \sys achieves higher accuracy than the single-model architectures. 
This is because, for example, GPT-4 performs well for schema normalization, not for verification.
Our dual-model architecture leverages the strengths of each model by assigning GPT-4 to schema generation and o1-mini to verification.

\noindent~\textbf{Impact of verification.}
We evaluate the impact of verification by comparing \sys with and without the verification module.
Table~\ref{tab:prompt-compare} shows the results.
We can clearly see that the verification module improves the accuracy for all normal forms.
This is because the feedback of the verification module makes the generation module refine the output schema, improving the quality of the output schema.

\begin{figure}[t!]
\centering
\subfloat[Number of required attempts]{\includegraphics[width=0.490\linewidth]{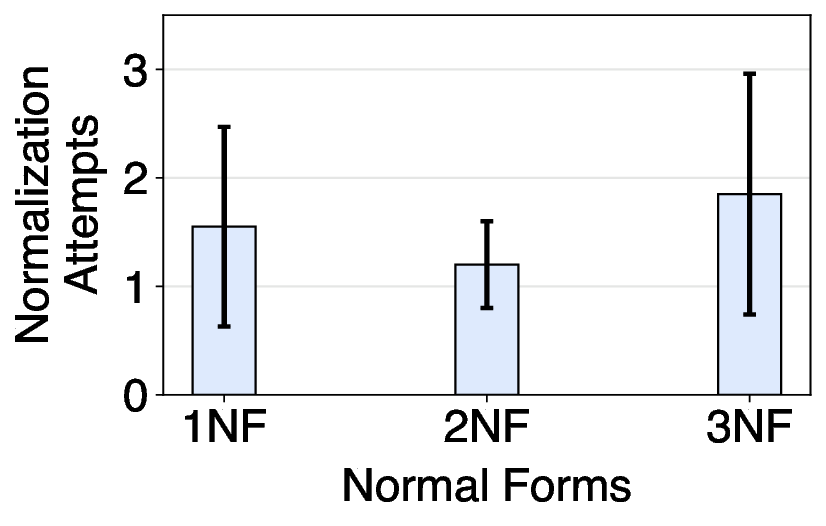}}
\subfloat[Accuracy]{\includegraphics[width=0.490\linewidth]{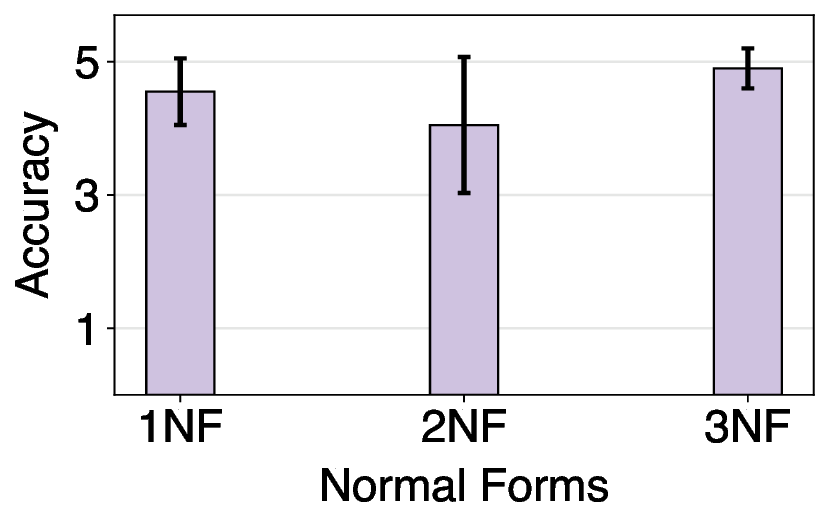}}
\caption{Impact of number of refinement loops. Most normalization tasks are completed within three iterations.
\label{fig:numofloops}}
\end{figure}

\noindent~\textbf{Impact of the number of refinement loops.}
In this experiment, we inspect the impact of the number of refinement loops. 
Fig.~\ref{fig:numofloops} and shows the average number of normalization attempts to finish the task and accuracy across normal forms in \sys.
We set the maximum refinement attempts to 20. 
We observe that most cases successfully converge within 3 attempts, and there are no cases where the number of attempts is more than 6.
We also observe that after 3 iterations, the LLM generally struggles to resolve remaining anomalies despite detailed feedback. 
Based on this result, we set the maximum number of self-refinement loops to 3, balancing high accuracy and cost efficiency.

\begin{figure}[t!]
\centering
\includegraphics[width=7.0cm]{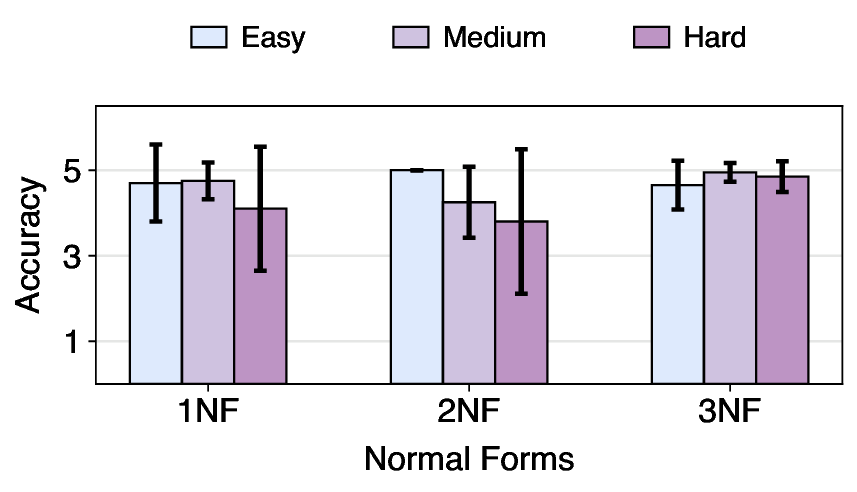}
\caption{Accuracy under different schema complexity.
\label{fig:normalization_complexity}}
\end{figure}

\noindent~\textbf{Accuracy under different schema complexity.}
We evaluate the normalization accuracy of \sys under different schema complexities using different target schemas shown in Table~\ref{tab:datasets}.
Fig.~\ref{fig:normalization_complexity} shows that \sys maintains consistently high accuracy for the \textit{Easy} and \textit{Medium} schemas.
However, as schema complexity increases to the \textit{Hard} level, normalization accuracy slightly decreases with larger standard deviations.
However, this is not a failure of the model but rather a reflection of the inherent semantic ambiguity present in complex, real-world schemas. In 'Hard' complexity scenarios, table relationships often allow for multiple valid normalization paths depending on domain-specific interpretations. The variance in our results indicates that Miffie is exploring these valid alternative interpretations rather than hallucinating incorrect schemas. Furthermore, despite the complexity, Miffie maintains a high success rate (approx. 4.85/5.0 for 3NF), proving its robustness in practical engineering environments where perfect consensus on schema design is rare.

\section{Conclusion \label{conclusion}}
We present \sys, a framework that automates database schema normalization using LLMs to minimize manual effort while maintaining high accuracy.
Built on a dual-model self-refinement architecture with task-specific zero-shot prompts, \sys effectively normalizes complex schemas and shows that dual-model refinement can outperform single-model approaches in domain-specific tasks.
In future work, we plan to extend our dual-model architecture to handle higher normal forms, such as Boyce-Codd Normal Form (BCNF) and Fourth Normal Form (4NF).
Since our framework is prompt-driven, this extension can be achieved by updating the verification module’s instructions to detect multi-valued dependencies and other advanced anomalies.

\appendices

\appendices 

\section{}
\subsection{Prompt Templates}
\label{app:prompts}
\begin{figure}[t!]
\centering
\includegraphics[width=0.8\linewidth]{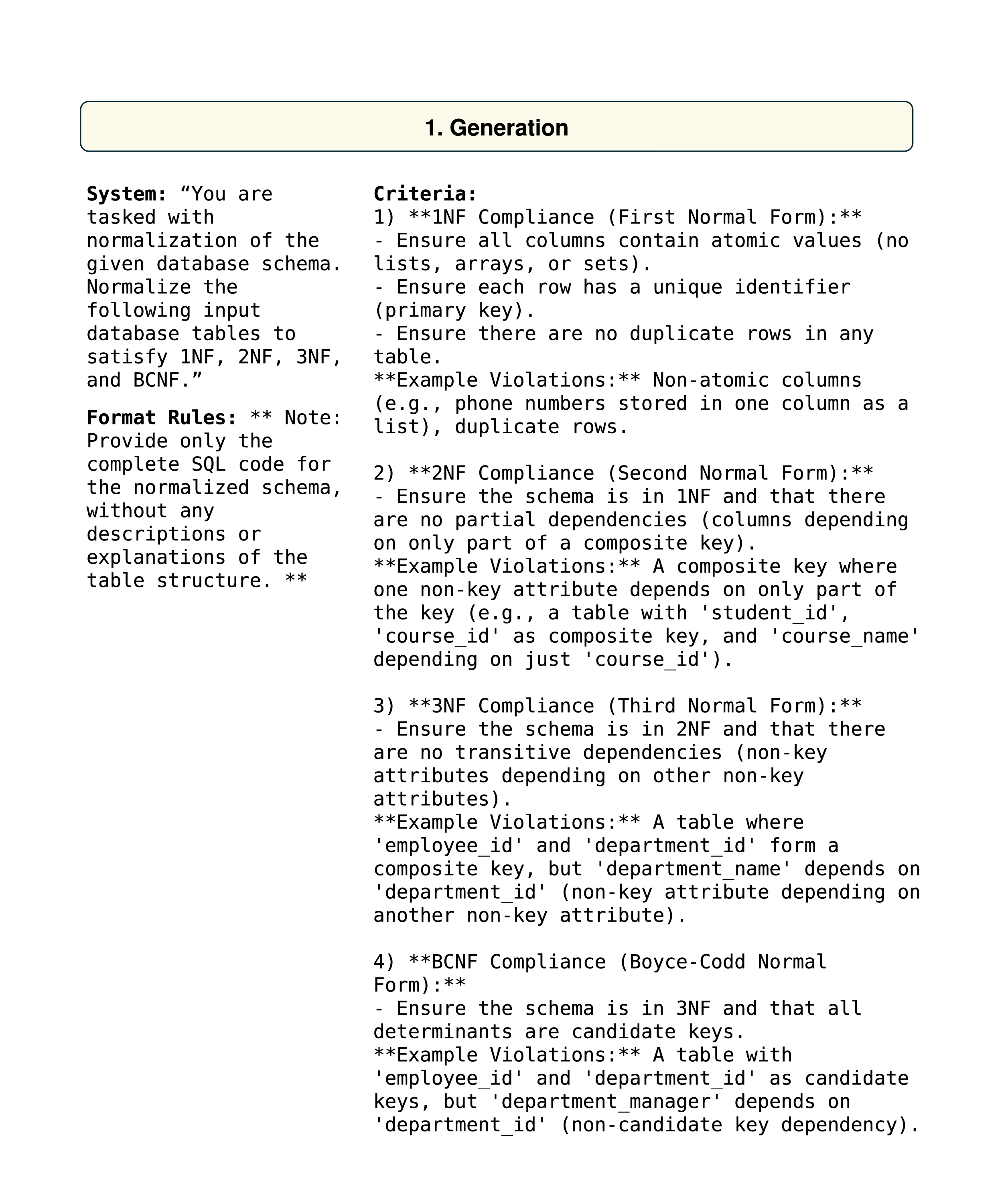}
\caption{Generation prompt for schema normalization.}
\label{fig:generation_prompt}
\end{figure}

\begin{figure}[t!]
\centering
\includegraphics[width=0.8\linewidth]{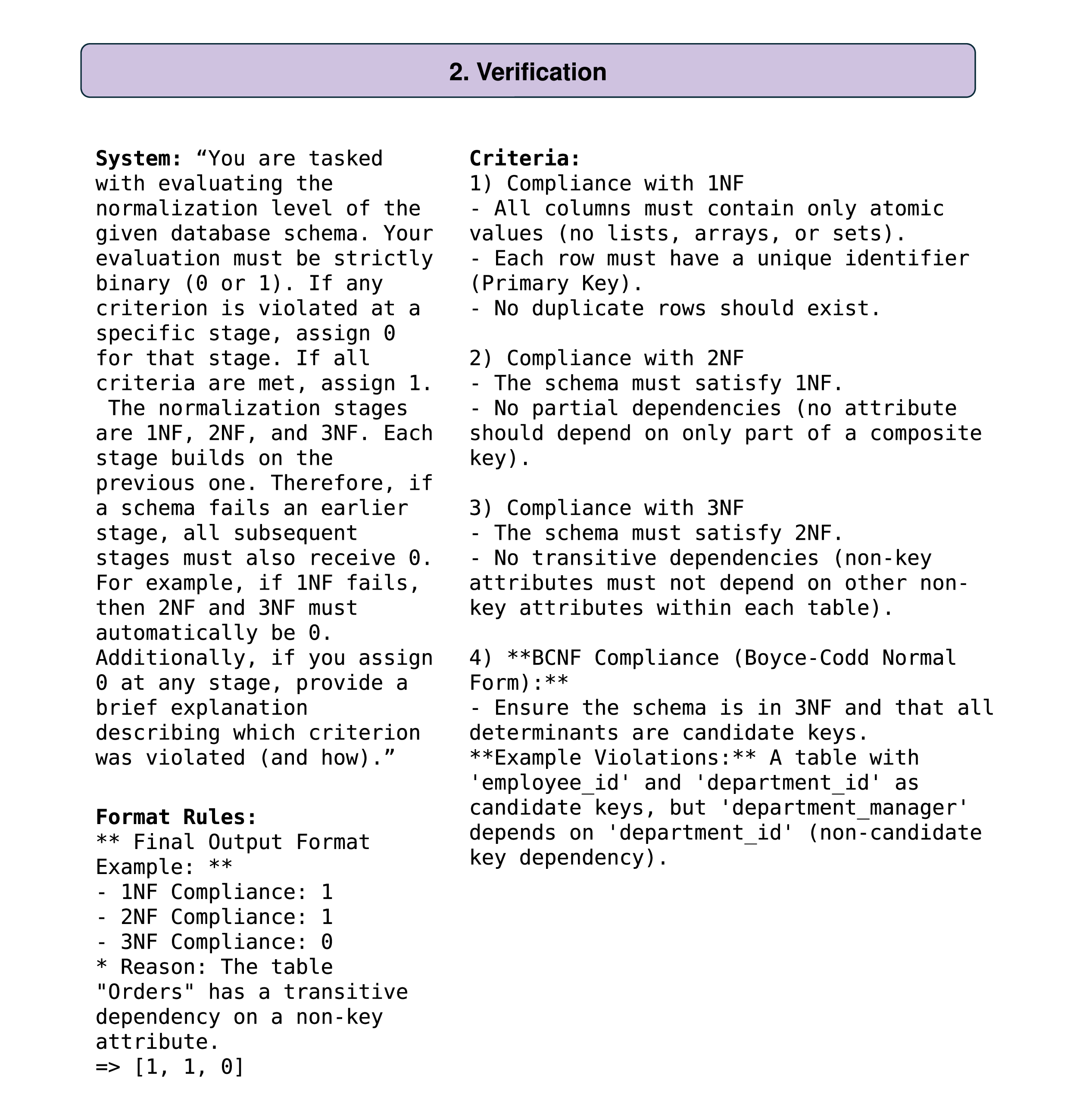}
\caption{Verification prompt for detecting normal-form violations.}
\label{fig:verification_prompt}
\end{figure}

We used the following prompts (Fig.~\ref{fig:generation_prompt} and Fig.~\ref{fig:verification_prompt}).

\FloatBarrier 

\subsection{Anomaly Injection Details}
\label{app:anomaly_injection}

This appendix describes how we synthetically inject anomalies into normalized target schemas to construct inputs that violate each normal form. 
For each stage (1NF, 2NF, and 3NF), we inject five anomalies by deliberately adding schema elements that contradict the corresponding definition.
Whenever possible, we distribute injected anomalies across different tables and relationships to avoid concentrating violations in a single relation. 

    
\begin{lstlisting}[style=sqlstyle,basicstyle=\ttfamily\scriptsize,caption={Example of 1NF anomaly injection in the \textit{Advertising} schema.},label={lst:adv_1nf_injection}]
CREATE TABLE Advert_Publishers (
    publisher_id INT PRIMARY KEY,
    publisher_details VARCHAR(255)
);

CREATE TABLE Adverts (
    advert_id INT PRIMARY KEY,
    publisher_id INT,
    publication_details VARCHAR(255),
    FOREIGN KEY (publisher_id) REFERENCES Advert_Publishers(publisher_id)
);

CREATE TABLE Tags (
    tag_word VARCHAR(255) PRIMARY KEY,
    tag_description VARCHAR(255)
);

CREATE TABLE Adverts_Tags (
    advert_id INT,
    tag_word VARCHAR(255),
    tags_list VARCHAR(255), -- list-like data violates 1NF
    PRIMARY KEY (advert_id, tag_word),
    FOREIGN KEY (advert_id) REFERENCES Adverts(advert_id),
    FOREIGN KEY (tag_word) REFERENCES Tags(tag_word)
);

CREATE TABLE Online_Users (
    user_id INT PRIMARY KEY,
    user_details VARCHAR(255),
    contact_numbers VARCHAR(255) -- comma-separated list violates 1NF
);

CREATE TABLE User_Interests (
    user_id INT,
    tag_word VARCHAR(255),
    PRIMARY KEY (user_id, tag_word),
    FOREIGN KEY (user_id) REFERENCES Online_Users(user_id),
    FOREIGN KEY (tag_word) REFERENCES Tags(tag_word)
);

CREATE TABLE Online_Users_Adverts (
    online_users_adverts_id INT PRIMARY KEY,
    advert_id INT,
    tag_word VARCHAR(255),
    user_id INT,
    date_and_time DATETIME,
    other_details VARCHAR(255),
    purchase_history VARCHAR(255), -- comma-separated list violates 1NF
    FOREIGN KEY (advert_id) REFERENCES Adverts(advert_id),
    FOREIGN KEY (tag_word) REFERENCES Tags(tag_word),
    FOREIGN KEY (user_id) REFERENCES Online_Users(user_id)
);
\end{lstlisting}

\noindent\textbf{1NF anomaly injection (atomicity violations).}
To violate 1NF, we introduce non-atomic attributes by encoding multivalued data as list-like strings (e.g., comma-separated values) in otherwise atomic columns. 
Listing~\ref{lst:adv_1nf_injection} shows a concrete example on the \textit{Advertising} schema.

\noindent\textbf{2NF anomaly (partial dependency on composite keys).}
To violate 2NF, we introduce \emph{partial dependencies} by (i) creating or selecting a relation with a \emph{composite primary key}, and (ii) adding a non-key attribute that depends on only a \emph{proper subset} of the composite key. 

\noindent\textbf{3NF anomaly injection (transitive dependencies).}
To violate 3NF, we inject \emph{transitive dependencies} among non-key attributes by introducing an intermediate attribute that functionally determines another non-key attribute.

\bibliographystyle{IEEEtran}
\bibliography{reference}

\end{document}